\documentclass[twocolumn]{aastex62}
\usepackage{times,graphics}
\usepackage{hyperref}

\usepackage{graphicx}
\usepackage{dcolumn}
\usepackage{bm}
\usepackage{natbib}
\usepackage{pstricks}
\usepackage{epsfig}
\usepackage{epstopdf}
\usepackage{multirow}
\usepackage{amsmath}
\usepackage{lipsum}

\begin{document}

	\title{Fluctuations in the Ginzburg-Landau Theory of Dark Energy:\\ internal (in-)consistencies in PLANCK data set}

	\author{Abdolali Banihashemi}
	\email{a\_banihashemi@sbu.ac.ir}
	\affiliation{Department of Physics, Shahid Beheshti University, G.C., Evin, Tehran 19839, Iran}
	
	\author{Nima Khosravi}
	\email{n-khosravi@sbu.ac.ir}
	\affiliation{Department of Physics, Shahid Beheshti University, G.C., Evin, Tehran 19839, Iran}

	\date{\today}

	\begin{abstract}

		In this work, predictions of the Ginzburg-Landau theory of dark energy (GLT) \citep{Banihashemi:2018has} for CMB lensing are studied. We find that the time and scale dependence of the dark energy fluctuations in this semi-phenomenological model is favored by data in several ways. Firstly, unlike $\Lambda$CDM, $\ell\leq801$ and $\ell>801$ ranges of the CMB angular power spectrum are consistent in this framework. Secondly, the lensing amplitude $A_L$ is completely consistent with unity when GLT is confronted with CMB data, even without including CMB lensing data. Therefore lensing anomaly is absent in this model. Although the background evolution of dark energy in this model is able to reconcile the $H_0$ inferred from CMB with that directly measured through observing nearby standard candles, the inclusion of BAO data brings the inferred $H_0$ close to what $\Lambda$CDM predicts and hence the Hubble tension is not fully eased. However, this doesn't affect the posterior on $A_L$ and the lensing anomaly is still absent.

	\end{abstract}

	\section{Introduction}
	
	Recently we have explored the framework of a phase transiting dark energy \citep{Banihashemi:2018has,Banihashemi:2018oxo,Banihashemi:2020wtb}. This idea states that dark energy as a self interacting many body system is sensitive to the cooling down of the universe and consequently undergoes a phase transition. To investigate this proposal, we have built a model, based on the literature of critical phenomena and more specifically Ginzburg-Landau theory \citep{Ginzburg:1950sr}. Therefore we have dubbed our model Ginzburg-Landau Theory of Dark Energy or just GLT for the sake of brevity. So far, we have mostly focused on GLT at the background level and showed that to some extent, we can address the famous $H_0$ tension\footnote{To have a complete review on this tension, cf. \citep{DiValentino:2021izs}.} via the dynamics that this model implies on dark energy.  However, producing some specific sort of dynamical dark energy is the least thing this model can predict. In fact, going beyond mean-field approximation and considering the spatial variations of dark energy, opens a rich area of phenomenology and at the same time makes the model more falsifiable. This paper is our first attempt to investigate GLT beyond mean field regime. We are particularly interested in the spatial features of our model since there are several spatial oddities in cosmological observations. For instance, it has been reported that there is an inconsistency between low and high multipoles of the CMB temperature angular power spectrum \citep{Aghanim:2018eyx}. According to what Planck team has revealed, when one put constraints on flat $\Lambda$CDM parameters, using $\ell \leq  801$ of the CMB temperature angular power spectrum, the resulted confidence regions are inconsistent with the case when they are constrained using the $\ell\geq802$ range. Also, they have shown that by allowing either $\Omega_k$ or the lensing amplitude, $A_L$, to vary as a free parameter, this discrepancy goes away. But none of these extensions to $\Lambda$CDM are really a solution for this issue. First, $A_L\neq1$ is not favored theoretically. It only gives a phenomenological hint about the cause of the problem. In fact, it introduces the so-called lensing anomaly which is related to low and high-$\ell$ inconsistency. Second, As again it's been asserted by Planck team, CMB data itself prefers a closed universe with nonzero $\Omega_k$ \citep{Aghanim:2018eyx}. But it's been shown that a nonzero spatial curvature can raise a crisis in cosmology \citep{DiValentino:2019qzk,Handley:2019tkm},	as the remaining cosmological observables are in strong disagreement with this curvature. There are several promising ideas for these problems to be alleviated. For example in \citep{Moshafi:2020rkq}, authors have shown that $\ddot{u}\Lambda$CDM model relaxes the low- and high-$\ell$'s inconsistency and the CMB lensing tension simultaneously, even better than $\Lambda$CDM+$A_L$ model. In \citep{Farhang:2020sij,Khosravi:2021csn}, it's been shown that a phenomenological gravitational phase transition can address these concerns too. In \citep{omena}, it's been shown that modifying the neutrino equation of state due to some unknown long-range interaction mimics $A_L>1$ and hence alleviates lensing anomaly.
	
	In this work, we are going to investigate the time and scale dependence of dark energy patches in the GLT framework to see if they can relax the lensing anomaly and hence low and high-$\ell$ inconsistency in Planck data set. In section \ref{GLT} this model  is reviewed and in section \ref{GLTlens}, the contribution of dark energy patches in the lensing potential power spectrum is studied. In section \ref{GLTRes} we explain the data sets we have used and also the results. Finally, we end this paper by a few concluding remarks.

	\section{Ginzburg-Landau Theory of Dark Energy}\label{GLT}
	In this model we assume that a scalar field, say $\phi$, is responsible for the dark energy evolution, in a way that
	\begin{equation}
	\Omega_\Lambda(z)=\langle\phi(\boldsymbol{r},z)\rangle,\label{Omega}
	\end{equation}
	where the average is over space; and the so-called Landau free energy governing this field is:
	\begin{equation}
	L=\int d^3\boldsymbol{r}\bigg[\frac{\gamma}{2}(\nabla\phi)^2+m\,t\,\phi^2+\frac{1}{2}\lambda\,\phi^4\bigg].\label{LFE}
	\end{equation}
	In the above expression $t$ is the reduced temperature, $t\equiv \frac{T-T_c}{T_c}$, where $T$, the temperature of dark energy, is assumed to be proportional to inverse of the scale factor, i.e. $T\propto(1+z)$, and $T_c$ is a critical temperature at which transition happens and $\langle\phi(\boldsymbol{r},z)\rangle$ takes a non zero value; and $\gamma$, $m$ and $\lambda$ are some constants that we try to put constraint on them. As it is evident in Eq. (\ref{LFE}), we have supposed $\mathbb{Z}_2$ symmetry for the Landau free energy; i.e. there isn't any odd power of $\phi$. This implies for the phase transition to be of second order and continuous; therefore the density changes smoothly in time. Also if we want to have an interaction term, we should have added a term like $H(\boldsymbol{r},z)\phi(\boldsymbol{r},z)$, where $H(\boldsymbol{r},z)$ represents all the external fields which might have interaction with $\phi(\boldsymbol{r},z)$.  The behavior of the field is governed by the equation of motion deduced by  demanding $L$ to be stationary with respect to variations of $\phi$:
	\begin{eqnarray}
	\frac{\delta L}{\delta\phi}=0\Longrightarrow
		-\gamma\nabla^2\phi+2mt\phi+2\lambda\phi^3=0.
	\end{eqnarray}
	If we consider $\phi$ within the mean field approach, $\phi(\boldsymbol{r},z)=\langle\phi(\boldsymbol{r},z)\rangle\equiv M(z)$, then it should obey the following equation:
	\begin{equation}
	2\,m\,t\,M+2\lambda\, M^3=0,
	\end{equation}
	 which implies:
	 \begin{equation}\label{BG}
	 M(t)=0\quad\text{or}\quad M(t)=\pm\sqrt{-\frac{mt}{\lambda}}.
	 \end{equation}
	The case $M=0$ corresponds to $T>T_c$ (or $t>0$). This means at $z>z_t$, the density of dark energy is effectively zero. The two other cases belong to the temperatures below $T_c$ (or $t<0$): after spontaneous symmetry breaking, $M$ takes one of these two possible values. We suppose $M$ takes the positive one\footnote{In fact, this means that our model consists of a sort of ``tuning", but not a ``fine" one; because the chance of happening in either cases is fifty percent.} and also assumption of spatial flatness of the universe, fixes the ratio $m/\lambda$ and we have:
	\begin{align}\label{zc}
	\Omega_\Lambda(z)=\Omega_\Lambda\sqrt{\frac{z_t-z}{z_t}},
	\end{align}	
	where $\Omega_\Lambda=1-\Omega_m-\Omega_r$ and $z_t$ are the fractional density of dark energy today and the critical redshift corresponding the transition temperature respectively. So our model, at the background level, has one extra free parameters more than flat $\Lambda$CDM: $z_t$; and the Friedman equation takes its new form:
	\begin{equation}
		H^2=H_0^2[\Omega_m (1+z)^3+\Omega_r(1+z)^4+\Omega_\Lambda(z)].
	\end{equation}
	This is almost all about GLT at the background level \citep{Banihashemi:2020wtb} and one can easily put constraints on the free parameters using background data which are mostly geometrical. But we are also interested in the predictions of this model when the ``fluctuations'' are also considered. In principle $\phi(\boldsymbol{r},z)$ can have spatial variations or ``patches of dark energy with different densities"\footnote{For more illustrations, cf. FIG. 4, 5 and 7 of \citep{Banihashemi:2018has}.}. This patchy pattern or fluctuations in the order parameter is very similar to different magnetic areas in a ferro-magnetic substance: each region has its own magnetic orientation and one can partition the substance into equi-magnetic areas. In addition to these fluctuations, there are also tiny ``perturbations'' on top of the average of the order parameter in every patch. In this work, we won't go into perturbations detail and only stick to fluctuations, as they have the dominant effect. These fluctuations, which their statistics are given by the Ginzburg-Landau theory, affect cosmological observables. Among them, the change in the CMB lensing pattern intrigued us the most and is studied here. Since these patches have different densities, they induce extra gravitational potential in the perturbed FLRW metric and the trajectories of photons are changed, accordingly. In the following sections, we try to investigate this effect. Before doing so, we will derive the power spectrum of the field $\phi$ which corresponds to the power spectrum of the dark energy density fluctuations. Basically, we are interested in two-point correlation function of the field $\phi$. Having the partition function of the system as \citep{golden}
	\begin{equation}
	\mathcal{Z}=\int\mathcal{D}\phi\, e^{-\beta L[\phi]},
	\end{equation}
where again, $L$ is the so-called Landau free energy, but this time with the external interaction term $H(\boldsymbol{r})\phi(\boldsymbol{r})$,
	\begin{equation}
L=\int d^3\boldsymbol{r}\bigg[\frac{\gamma}{2}(\nabla\phi(\boldsymbol{r}))^2+mt\phi^2(\boldsymbol{r})+\frac{1}{2}\lambda\phi^4(\boldsymbol{r})-H(\boldsymbol{r})\phi(\boldsymbol{r})\bigg].\label{LFE2}
\end{equation}
It easily follows that the average of the field or its one-point function is obtained as: 
\begin{equation}
\langle\phi(\boldsymbol{r})\rangle=\frac{1}{\beta}\,\frac{\delta\ln\mathcal{Z}}{\delta H(\boldsymbol{r})},
\end{equation}
and the two-point function can be extracted as
\begin{align}
G(\boldsymbol{r}-\boldsymbol{r}')&\equiv\langle\phi(\boldsymbol{r})\phi(\boldsymbol{r}')\rangle-\langle\phi(\boldsymbol{r})\rangle\langle\phi(\boldsymbol{r}')\rangle\nonumber
\\
&=\frac{1}{\beta^2}\,\frac{\delta^2\ln\mathcal{Z}}{\delta H(\boldsymbol{r}')\,\delta H(\boldsymbol{r})}.
\end{align}
In the above expressions, $\beta=1/T$ and by $\delta$ we mean functional derivative. In order to obtain an equation for $G(\boldsymbol{r}-\boldsymbol{r}')$, we can make use of the equation of motion for $\phi(\boldsymbol{r})$ and variate it with respect to $H(\boldsymbol{r}')$:
\begin{align}
&-\gamma\nabla^2\phi(\boldsymbol{r})+2mt\phi(\boldsymbol{r})+2\lambda\phi^3(\boldsymbol{r})-H(\boldsymbol{r})=0
\\
&\xrightarrow{\delta/\delta H(\boldsymbol{r}')}\nonumber
\\
&\beta\,[-\gamma\nabla^2+2mt+6\lambda\phi^2(\boldsymbol{r})]\,G(\boldsymbol{r}-\boldsymbol{r}')=\delta(\boldsymbol{r}-\boldsymbol{r}').
\end{align}
This equation governs the two-point correlation function of $\phi$. By going to Fourier space, not only it becomes easier to solve it, but also we directly arrive at the power spectrum which was desired. Depending on being above $T_c$ or below it, $\phi$ follows one of the behaviors mentioned in Eq. (\ref{BG}) and for the power spectrum we have:  
\begin{equation}
\label{phipower} \mathcal{P}_\phi(k,z)=\frac{1+z}{\gamma}\frac{1}{k^2+\xi^{-2}(z)},
\end{equation}	
	where the correlation length, $\xi(z)$, in any of the above or below $T_c$ regimes, is defined as:

\begin{align}
	\label{corrlength} \xi(z)\equiv\bigg(\frac{\gamma\,(1+z_t)}{2\,m\,(z-z_t)}\bigg)^{1/2}\quad\text{for}\quad z>z_t,\nonumber
	\\
	\xi(z)\equiv\bigg(\frac{\gamma\,(1+z_t)}{4\,m\,(z_t-z)}\bigg)^{1/2}\quad\text{for}\quad z<z_t.
\end{align}

	\section{CMB lensing in the GLT}\label{GLTlens}
	In order to compute the lensing potential power spectrum, firstly we need power spectra of Bardeen potentials. Bardeen potentials are related to the density fluctuations through the Poisson equation, which in Fourier space reads as follows: 
	\begin{align}
\label{Poisson}	k^2\Psi&=4\pi G (1+z)^{-2}(\delta\rho_m+\delta\rho_{de})\nonumber
\\
&=4\pi G(1+z)^{-2}\bigg[\bar{\rho}_m\delta_m+\frac{3H_0^2}{8\pi G}\delta\phi\bigg]\nonumber
\\
&=4\pi G(1+z)^{-2}\frac{3H_0^2}{8\pi G}\bigg[\Omega_m(1+z)^3\delta_m+\delta\phi\bigg]\nonumber
\\
&=\frac{3}{2}H_0^2\bigg[\Omega_m(1+z)\delta_m+\frac{\delta\phi}{(1+z)^2}\bigg].
	\end{align}
	So for two-point correlation functions we have:
	\begin{align}
\label{2point} \langle\Psi\Psi\rangle=\frac{9\ H_0^4}{4\ k^4}\bigg[\Omega_m^2(1+z)^2\langle\delta_m\delta_m\rangle+\frac{\langle\delta\phi\delta\phi\rangle}{(1+z)^4}\bigg],
\end{align}
therefore 
\begin{align}
\label{power}\mathcal{P}_\Psi(k;z,z')=\frac{9\ H_0^4}{4\ k^4}\bigg[\Omega_m^2\mathcal{P}_{\delta_m}(k,z=\infty)(1+z)T_{\delta_m}(k,z)\nonumber
\\
 \times(1+z')T_{\delta_m}(k,z')+\frac{\sqrt{\mathcal{P}_{\phi}(k,z)}}{(1+z)^2}\frac{\sqrt{\mathcal{P}_{\phi}(k,z')}}{(1+z')^2}\bigg],
	\end{align} 
	where $\mathcal{P}$ and $T_{\delta_m}$ are power spectrum and matter transfer function respectively.
In the above calculation, we assumed that matter and dark energy fluctuation fields are independent and ignore the cross term. The power spectrum of dark energy fluctuations is as described in Eqs. (\ref{phipower}) and (\ref{corrlength}). Therefore, the lensing potential power spectrum reads:
\begin{align}
C_\ell^{\psi_{tot}}=C_\ell^{\psi_m}+C_\ell^{\psi_{de}},
\end{align}
where
\begin{eqnarray}
C_\ell^{\psi_m}&\equiv&36\,\pi\, H_0^4\,\Omega_m^2\,\int\frac{dk}{k^5}\,\mathcal{P}_{\delta_m}(k,z=\infty)\nonumber
\\
&\times&\bigg[\int_{0}^{z_d}\frac{dz}{H(z)}(1+z)T_{\delta_m}(k,z)j_\ell(k\chi)\bigg(\frac{\chi-\chi_*}{\chi\chi_*}\bigg)\bigg]^2,\nonumber
\end{eqnarray}
	and
\begin{eqnarray}
	C_\ell^{\psi_{de}}&\equiv&36\,\pi\, H_0^4\,\int\frac{dk}{k^5}\hspace{3.2cm}\nonumber
	\\
	&\times&\bigg[\int_{0}^{z_d}\frac{dz}{H(z)(1+z)^2}\sqrt{\mathcal{P}_{\phi}(k,z)}\ j_\ell(k\chi)\bigg(\frac{\chi-\chi_*}{\chi\chi_*}\bigg)\bigg]^2.\nonumber
\end{eqnarray}
So our model predicts an extra term for lensing potential power, namely $C_l^{\psi_{de}}$. In the following section we try to see if this extra term has the features desired by data or not. For this purpose, we implemented our model into the publicly available code, \texttt{CAMB} \citep{Lewis:1999bs}, to calculate the two-point statistics of the cosmic background radiation anisotropies. We did so by modifying both expansion history of the universe and source of the Poisson equation, by adding the fluctuation amplitude of dark energy in each scale and time, $k,z$, as a new source for the gravitational potential. Since there isn't any non-gravitational interaction between dark energy and other other fluids, we left the Boltzmann’s equations as they  are in LCDM model. To sample the parameter space and find the confidence regions of the free parameters, we made use of \texttt{CosmoMC} \citep{Lewis:2002ah,Lewis:2013hha}. In this work, the threshold for the chains convergence measure, $R-1$, was set to $0.025$. Our statistical analysis and plotting are done by using \texttt{Getdist} \citep{Lewis:2019xzd}.

\section{Confronting with data and results}\label{GLTRes}

For our purposes, we first split the CMB's TT power spectrum into low- and high-$\ell$'s and check our model against them separately. This allows us to check if they are consistent in the GLT framework or not. If they are consistent then we are allowed to work with full CMB power spectrum data.

\begin{figure}[h]
	\centering
	\includegraphics[width=\linewidth]{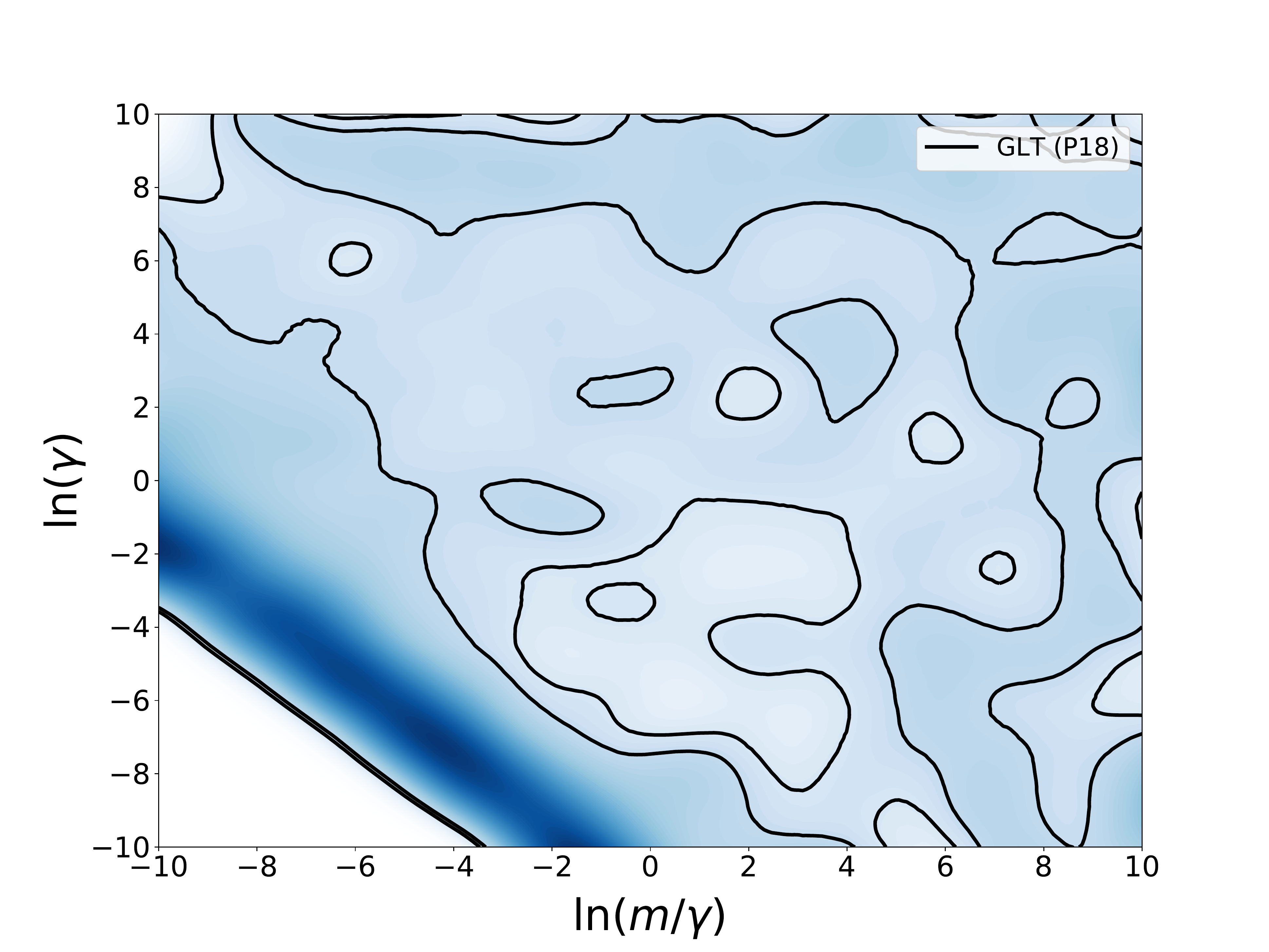}
	\caption{In this plot, two dimensional likelihood of $\ln(\gamma)$ vs. $\ln(m/\gamma)$, is shown; where P18 is used. Darker colors correspond to higher confidence levels. As it can be seen, there is a line of high confidence. We infer the equation of this line as $\ln(m/\gamma)=-\ln(\gamma)-12$ and hence we no longer keep $m$ as a free parameter.}
	\label{fig:m-gamma}
\end{figure}

Here are the data combinations we have used:
\begin{itemize}
	\item Planck 2018 CMB temperature power spectrum when $\ell$ ranges from $2$ to $801$. In addition, we have used CMB low-$\ell$ polarization power spectrum, i.e. \texttt{SimALL}. We refer to this combination as low-$\ell$ \citep{Aghanim:2019ame}.
	
	\item  Planck 2018 CMB temperature power spectrum when $\ell$ ranges from $802$ to $2500$. We have also added \texttt{SimALL} to this combination because without which optical depth, $\tau$, won't get constrained well. We refer to this combination as high-$\ell$ \citep{Aghanim:2019ame}.
	
	\item Full Plank 2018 temperature and polarization power spectra and their cross correlations. We refer to this combination as P18 \citep{Aghanim:2019ame}
	
	\item  P18, plus power spectrum of CMB lensing potential inferred from the four-point function of CMB temperature map. We refer to this combination as P18+lensing \citep{Planck:2018lbu}.
	
	\item P18, plus BAO volume distance measurements, at $z=0.32$ (LOWZ) \citep{Anderson:2013zyy}, $z=0.57$ (CMASS) \citep{Anderson:2013zyy}, $z=0.106$ (6dFGS) \citep{6df} and $z=0.15$ (MGS) \citep{Ross:2014qpa}. Furthermore we used BAO angular diameter distance measurements at $z=0.44$, $z=0.60$ and $z=0.73$ (WiggleZ) \citep{wigglez}. We refer to this data combination as P18+BAO.
\end{itemize}
The set of parameters on its members we would like to put constraint on is\footnote{We don't bring $m$ here because our very first analysis showed that CMB data fixes $m$ and the ratio of $m/\gamma$, present in the power spectrum, is able to vary only through variations of $\gamma$. This fact is shown and explained in Fig. \ref{fig:m-gamma}. For this reason, we have eliminated $m$ from our analysis.}:
\begin{equation}
\mathcal{P}=\{\Omega_bh^2,\Omega_ch^2,100\Theta_{MC},\tau,n_s,\ln[10^{10}A_s],a_t,\ln(\gamma),A_L\}.
\end{equation}

The first six parameters are common with $\Lambda$CDM and $A_L$ can be thought of as an extension to it. But $a_t$ and $\ln(\gamma)$ are two extra free parameters in GLT. In our analysis, we prefer to work with $a_t\equiv\frac{1}{1+z_t}$, instead of $z_t$, since equations and functions in \texttt{CAMB} are written in terms of scale factor.

In table \ref{bestfit}, the details about priors on these parameters and the data combination we have used for each model can be found. 

In FIG. \ref{fig:low-high-full-ell}, we have shown that the 1-$\sigma$ contours of low- and high-$\ell$'s are overlapped. This means they are consistent with each other in the GLT framework unlike $\Lambda$CDM. Hence it's statistically meaningful to combine these two data sets in the framework of GLT. 

\begin{figure}[h]
	\centering
	\includegraphics[width=\linewidth]{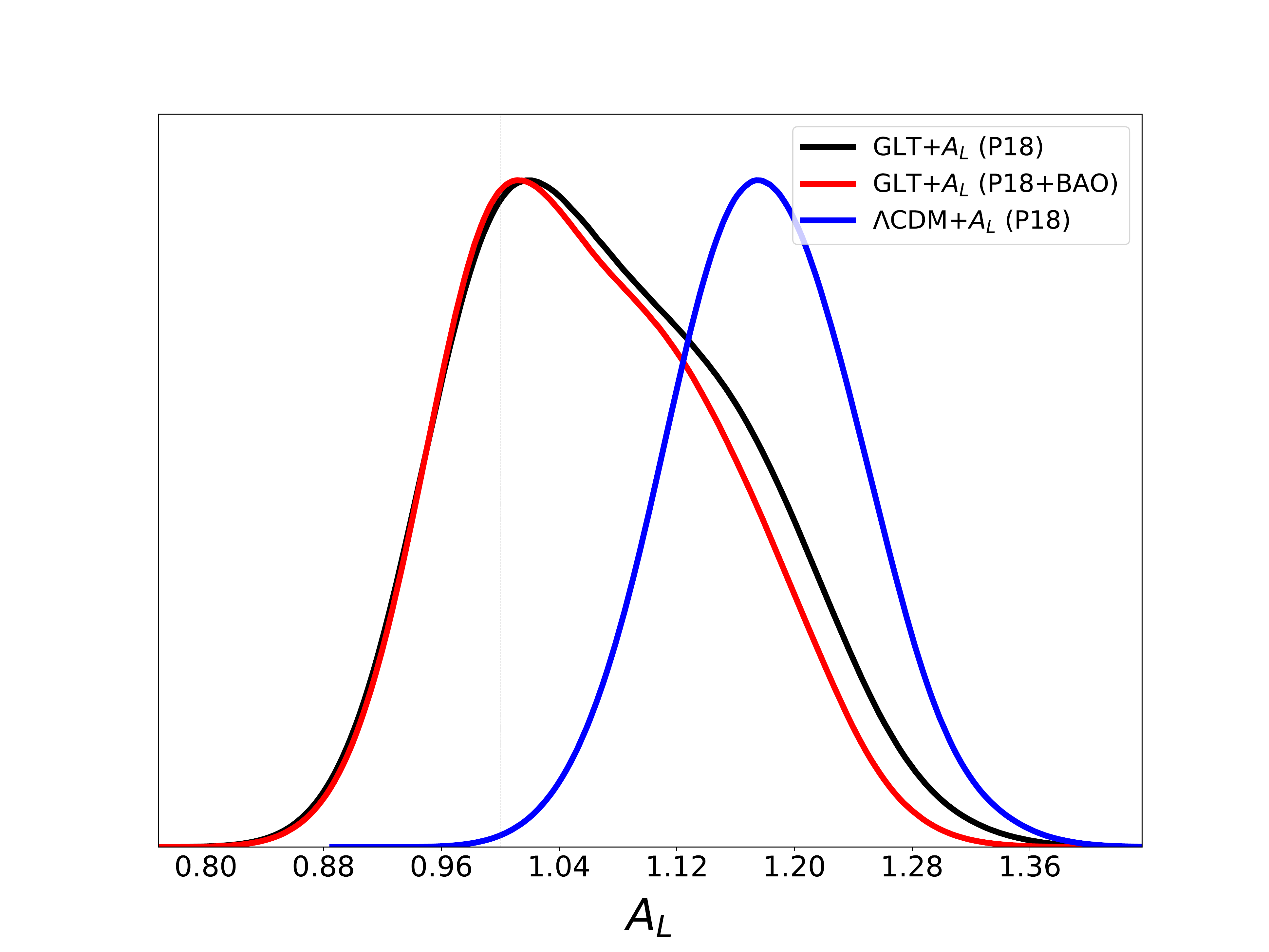}
	\caption{After making sure that low-$\ell$ and high-$\ell$ data sets are consistent within the framework of GLT, we combine these two datasets plus CMB polarization to see if $A_L=1$ is included in the high confidence range or not. To this end, we relaxed the $A_L$ and as it is evident in the black curve, $A_L=1$ is inside the 1-$\sigma$ confidence level. However, the blue curve indicates that $\Lambda$CDM is inconsistent with $A_L=1$ and technically it is rather invalid to combine P18 and lensing data sets within $\Lambda$CDM. The red curve shows that although BAO data makes $H_0$ decrease (c.f. last column of table \ref{bestfit}), it doesn't affect the posterior on $A_L$ and hence the main claim of this paper is still relevant.}
	\label{fig:al}
\end{figure}

Now it is time to consider the CMB-lensing anomaly. In the literature, as is also mentioned by \citep{Aghanim:2018eyx}, the CMB-lensing and low/high-$\ell$ inconsistencies are suggested to be related theoretically. The reason is that it is expected that the lensing potential affects high-$\ell$'s (small scales) more than low-$\ell$'s (large scales) in $C^\ell_{\rm TT}$'s. To check this proposal, the lensing amplitude $A_L$ became relaxed to see if our GLT model can relieve the CMB lensing anomaly. In FIG. \ref{fig:al}, we have shown that the fluctuations in GLT can indeed address the CMB-lensing anomaly completely. More quantitatively, in table \ref{bestfit}, it can be seen that while in $\Lambda$CDM+$A_L$ model, the $A_L=1$ is $2.8\sigma$ away from the best fit, in GLT+$A_L$, it is within $1\sigma$ confidence level. This is important because lensing data itself prefers $A_L\approx1$ and its inclusion pushes $A_L$ toward unity by force and raises the $\chi^2$ significantly. Note that this implicitly suggests that it is not statistically valid to combine P18 and lensing data in $\Lambda$CDM. In GLT though, $A_L=1$ is already accessible and the addition of lensing data doesn't penalize the model that much.  

\begin{figure}[h]
	\centering
	\includegraphics[width=\linewidth]{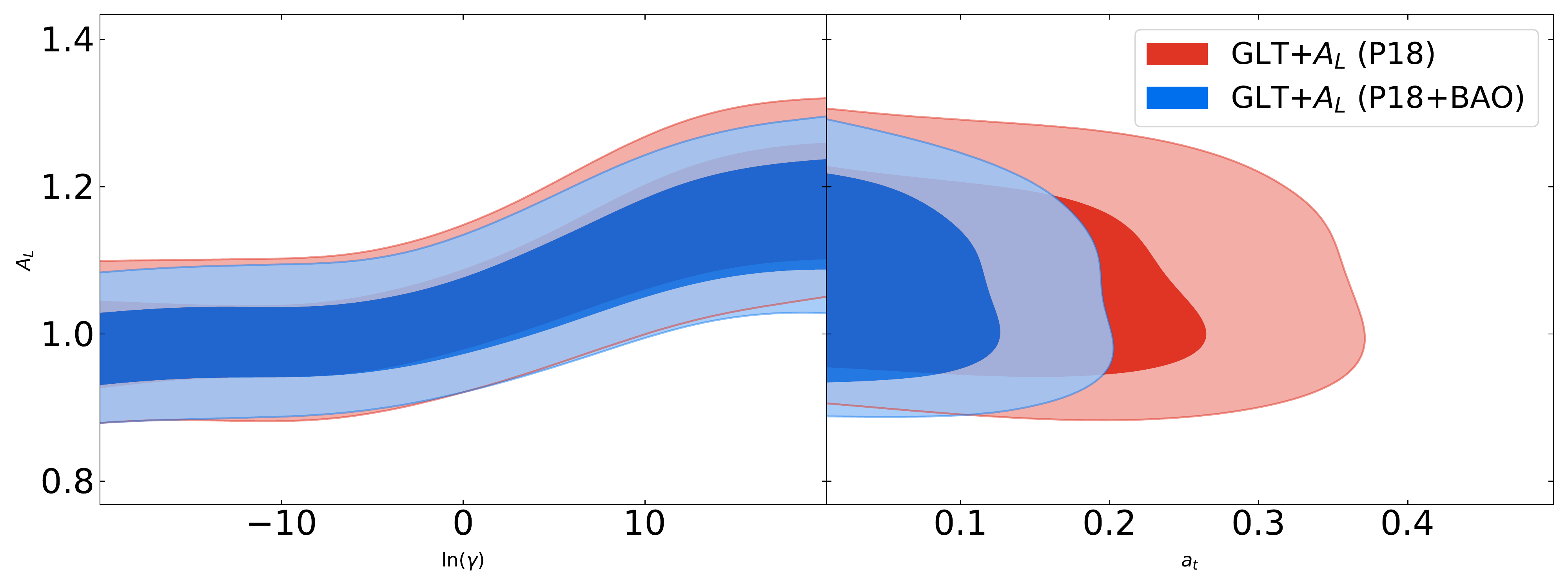}
	\caption{1-$\sigma$ and 2-$\sigma$ 2D plots of $A_L$ vs  $\ln(\gamma)$ and $a_t$. As can be seen,  $A_L$ is positively correlated to $\ln(\gamma)$; while it is insensitive to $a_t$. So if inclusion of BAO data push $a_t$ toward zero, $A_L=1$ will be still in the 1-$\sigma$ region of confidence.}
	\label{fig:rectangle}
\end{figure}

\begin{figure*}
	\centering
	\includegraphics[width=\linewidth]{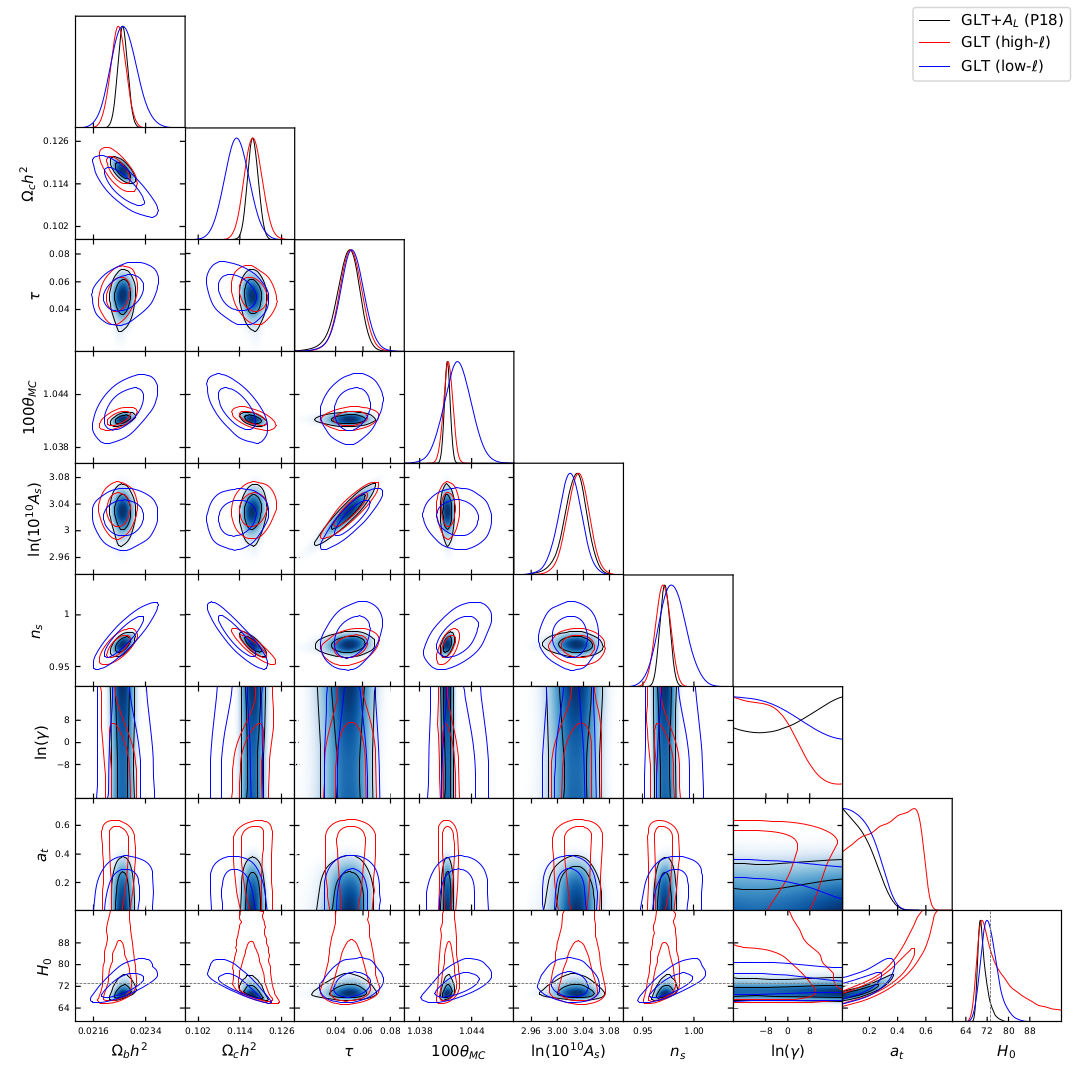}
	\caption{1-$\sigma$ and 2-$\sigma$ likelihoods for low-$\ell$ and high-$\ell$  CMB TT power spectrums are shown in blue and red, respectively. It is obvious in our GLT model there is no inconsistencies between these two datasets, in contrast to $\Lambda$CDM. This allows us to use the full CMB power spectrum (TTTEEE) to constrain the free parameters in the model. 1-$\sigma$ and 2-$\sigma$ contours are shown in shaded black for this case. Note that for the full P18 case we have allowed $A_L$ to vary. As it is shown in FIG.\ref{fig:al}, it is consistent with $A_L=1$.}
	\label{fig:low-high-full-ell}
\end{figure*}

On the other hand, the temporal evolution of the dark energy density that is described in Eq. (\ref{zc}), is able to ease the discrepancy between the derived $H_0$ from CMB and local measurements which are higher. Note that if we consider BAO's then there is no chance to address the $H_0$ tension but still we address the internal inconsistencies, as can be seen in FIG. \ref{fig:al} and FIG. \ref{fig:rectangle}.

\begin{table*}
	\begin{tiny}
		\begin{tabular}{|c|c|c|c|c|c|c|c|c|}
			
			\hline&&&&&&&&\\
			$\quad$Parameter$\quad$&$\quad$Prior$\quad$&$\Lambda$CDM (P18)&GLT (P18)&$\Lambda$CDM+$A_L$ (P18)&GLT+$A_L$ (P18)&$\Lambda$CDM+$A_L$ (P18+lensing)&GLT+$A_L$ (P18+lensing)&GLT+$A_L$ (P18+BAO)\\
			\hline&&&&&&&&\\
			$\Omega_bh^2$&[0.005\ ,\ 0.1]&$0.02236\pm 0.00015$&$0.02259\pm 0.00016$&$0.02259\pm 0.00017$&$0.02261\pm 0.00017$&$0.02248\pm 0.00016$&$0.02260\pm 0.00017$&$0.02252\pm 0.00015$\\&&&&&&&&\\
			
			$\Omega_ch^2$&[0.001\ ,\ 0.99]&$0.1202\pm 0.0014$&$0.1179\pm 0.0015$&$0.1181\pm 0.0016$&$0.1178\pm 0.0016$&$0.1185\pm 0.0015$&$0.1178\pm 0.0016$&$0.1188\pm 0.0011$\\&&&&&&&&\\
			
			100$\Theta_{MC}$&[0.5\ ,\ 10]&$1.04090\pm 0.00031$&$1.04119\pm 0.00032$&$1.04114\pm 0.00032$&$1.04119\pm 0.00033$&$1.04112\pm 0.00032$&$1.04124\pm 0.00033$&$1.04108\pm 0.00030$\\&&&&&&&&\\
			
			$\tau$&[0.01\ ,\ 0.8]&$0.0544^{+0.0070}_{-0.0081}$&$0.0500^{+0.0082}_{-0.0071}$&$0.0492^{+0.0088}_{-0.0073}$&$0.0487^{+0.0092}_{-0.0072}$&$0.0486^{+0.0090}_{-0.0076}$&$0.0492^{+0.0088}_{-0.0075}$&$0.0491\pm 0.0082$\\&&&&&&&&\\
			
			$n_s$&[0.8\ ,\ 1.2]&$0.9649\pm 0.0044$&$0.9712\pm 0.0048$&$0.9708\pm 0.0048$&$0.9716\pm 0.0049$&$0.9692\pm 0.0049$&$0.9716\pm 0.0049$&$0.9691\pm 0.0041$\\&&&&&&&&\\
			
			$\ln[10^{10}A_s]$&[2\ ,\ 4]&$3.045\pm 0.016$&$3.030^{+0.017}_{-0.015}$&$3.029^{+0.018}_{-0.015}$&$3.027^{+0.019}_{-0.016}$&$3.028^{+0.019}_{-0.016}$&$3.028^{+0.018}_{-0.016}$&$3.030^{+0.018}_{-0.016}$\\&&&&&&&&\\
			
			$a_t$&[0.01\ ,\ 1]&-&$< 0.329$&-&$< 0.189$&-&$< 0.190$&$< 0.171$\\&&&&&&&&\\
			
			$\ln(\gamma)$&[-16\ ,\ 16]&-&$-6.8^{+4.6}_{-13}$&-&not constrained&-&$< -2.65$&not constrained\\&&&&&&&&\\
			
			$A_L$&[0.5\ ,\ 2]&-&-&$1.180\pm 0.065$&$\boldsymbol{1.072^{+0.084}_{-0.11}}$&$1.051^{+0.037}_{-0.041}$&$\boldsymbol{1.002^{+0.038}_{-0.044}}$&$\boldsymbol{1.062^{+0.078}_{-0.11}}$\\&&&&&&&&\\
			
			$H_0\ \tiny{[\rm km/s/Mpc]}$&-&$67.27\pm 0.60$&$70.42^{+0.94}_{-2.4}$&$68.28\pm 0.72$&$70.41^{+0.96}_{-2.2}$&$68.05\pm 0.70$&$70.44^{+0.93}_{-2.3}$&$68.69^{+0.55}_{-0.68}$\\\hline&&&&&&&&\\
			
			Total $\chi^2_{\rm min}$&-&2772.6&$\boldsymbol{2764.8}$&2764.1&2763.5&$\quad2789.4\quad$&$\quad\boldsymbol{2780.0}\quad$&$2768.1
			$\\&&&&&&&&\\
			
			$\chi^2_{\rm P18}$&-&2768.9&$\boldsymbol{2759.6}$&2760.8&2760.5&$2777.4$&$\boldsymbol{2769.1}$&$2761$\\&&&&&&&&\\
			
			$\chi^2_{\rm lensing}$&-&-&-&-&-&9.2&$\boldsymbol{7.8}$&-\\&&&&&&&&\\
			
			$\chi^2_{\rm BAO}$&-&-&-&-&-&-&-&$4.6$\\&&&&&&&&\\
			\hline
		\end{tabular}	
	\end{tiny}
	
	\caption{Parameters versus models. The confidence regions here, are 68$\%$. One would infer that $\Lambda$CDM is not consistent with its own prediction about the amplitude of lensing unless we include lensing data. GLT, on the other hand, is able to have $A_L=1$ even when lensing data is not included. The last four raws, compares the models with the measure of $\chi^2$. As it can be seen, our model is favored over $\Lambda$CDM when confronting P18 alone. This is because the lensing effect is naturally enhanced and at high $\ell$'s theory has a better fit to data. When we add $A_L$ to the models, fluctuations of dark energy are not as useful as before and $\Lambda$CDM and GLT have more or less the same goodness of fit to P18. However, this goodness for $\Lambda$CDM comes at the cost of $A_L\neq1$, while GLT provides the possibility of being unity for $A_L$. This would get more apparent by adding lensing data since lensing data prefers $A_L=1$. In the last column, while the BAO data has made the $H_0$ to get close to what $\Lambda$CDM predicts, $A_L=1$ has remained inside the $1-\sigma$ confidence region. The absent raw above is $\chi^2_{\rm prior}$, which \texttt{CosmoMC} reports, but we didn't include it in the table.} \label{bestfit}
\end{table*}

\section{Concluding remarks and future perspective}\label{con}
	
	In this work, we studied the Ginzburg-Landau theory of dark energy (GLT) beyond its mean field approximation. The GLT is based on several assumptions: DE is somehow sensitive to the cooling down of the universe and hence undergoes a phase transition, the evolution of DE comes from the so-called Landau free energy, and any details about both temporal or spatial features of this DE can be deduced from this effective free energy. While in previous works \citep{Banihashemi:2018has,Banihashemi:2018oxo,Banihashemi:2020wtb}, we studied GLT at the background level without quantitative considerations of its spatial fluctuations, this work is our first attempt to see the fingerprints of GLT in its very specific spatial properties. This can be seen as the smoking gun for our GLT model since it is very different from other DE models e.g. quintessence. It turned out that the time and scale dependencies of the DE patches in this scenario are such that ease the low and high-$\ell$ inconsistency in CMB angular power spectrum (it can be checked qualitatively in FIG.\ref{fig:low-high-full-ell}) and also solves the CMB lensing anomaly (cf. FIG. \ref{fig:al} and table \ref{bestfit} for discussion). The addition of BAO data prevents GLT from fully solving the $H_0$ tension, but we would like to emphasize that while GLT fluctuations can justify the internal inconsistencies in Planck data and also solve the lensing anomaly, they won't worsen the Hubble tension at all. Note that relaxing $\Omega_k$ in $\Lambda$CDM framework does the same job, but on the other hand extremely intensifies the $H_0$ tension and also makes the BAO in strong disagreement with the predictions \citep{DiValentino:2019qzk}.
	
	There are also other peculiar features in GLT that lie beyond the scope of this work. As an example, this model predicts a transient longwave mode in dark energy density and this may challenge the isotropy of the universe. 
	\\
	In the future, we not only plan to search for other observational implications of the GLT, but also we wish to build it on a more concrete theoretical ground with less phenomenological bases.

	\acknowledgments
	AB is in debt to  the Institute for Theoretical Physics at the University of Heidelberg, where this work was initiated, for their hospitality. We would like to thank Luca Amendola for his insightful suggestions and comments during our talks on GLT.  We thank Iran National Science Foundation (INSF), for supporting this work under project no. 98022568.

	\newpage

	\bibliography{references}

\begin{thebibliography}{}
\expandafter\ifx\csname natexlab\endcsname\relax\def\natexlab#1{#1}\fi
\bibitem[Banihashemi et al.(2019)]{Banihashemi:2018has} Banihashemi, A., Khosravi, N., \& Shirazi, A.~H.\ 2019, \prd, 99, 083509. doi:10.1103/PhysRevD.99.083509

\bibitem[Banihashemi et al.(2020)]{Banihashemi:2018oxo} Banihashemi, A., Khosravi, N., \& Shirazi, A.~H.\ 2020, \prd, 101, 123521. doi:10.1103/PhysRevD.101.123521

\bibitem[Banihashemi et al.(2021)]{Banihashemi:2020wtb} Banihashemi, A., Khosravi, N., \& Shafieloo, A.\ 2021, \jcap, 2021, 003. doi:10.1088/1475-7516/2021/06/003

\bibitem[Ginzburg et al.(1950)]{Ginzburg:1950sr} Ginzburg, V., Landau, L.\ 1950, Zh. Eksp. Teor. Fiz. \textbf{20}, 1064-1082

\bibitem[Di Valentino et al.(2021)]{DiValentino:2021izs} Di Valentino, E., Mena, O., Pan, S., et al.\ 2021, Classical and Quantum Gravity, 38, 153001. doi:10.1088/1361-6382/ac086d

\bibitem[Planck Collaboration et al.(2020)]{Aghanim:2018eyx} Planck Collaboration, Aghanim, N., Akrami, Y., et al.\ 2020, \aap, 641, A6. doi:10.1051/0004-6361/201833910

\bibitem[Di Valentino et al.(2020)]{DiValentino:2019qzk} Di Valentino, E., Melchiorri, A., \& Silk, J.\ 2020, Nature Astronomy, 4, 196. doi:10.1038/s41550-019-0906-9

\bibitem[Handley(2021)]{Handley:2019tkm} Handley, W.\ 2021, \prd, 103, L041301. doi:10.1103/PhysRevD.103.L041301

\bibitem[Moshafi et al.(2021)]{Moshafi:2020rkq} Moshafi, H., Baghram, S., \& Khosravi, N.\ 2021, \prd, 104, 063506. doi:10.1103/PhysRevD.104.063506

\bibitem[Farhang \& Khosravi(2020)]{Farhang:2020sij} Farhang, M. \& Khosravi, N.\ 2020, arXiv:2011.08050

\bibitem[Khosravi \& Farhang(2021)]{Khosravi:2021csn} Khosravi, N. \& Farhang, M.\ 2021, arXiv:2109.10725

\bibitem[Lewis et al.(2000)]{Lewis:1999bs} Lewis, A., Challinor, A., \& Lasenby, A.\ 2000, \apj, 538, 473. doi:10.1086/309179

\bibitem[Lewis \& Bridle(2002)]{Lewis:2002ah} Lewis, A. \& Bridle, S.\ 2002, \prd, 66, 103511. doi:10.1103/PhysRevD.66.103511

\bibitem[Lewis(2013)]{Lewis:2013hha} Lewis, A.\ 2013, \prd, 87, 103529. doi:10.1103/PhysRevD.87.103529

\bibitem[Lewis(2019)]{Lewis:2019xzd} Lewis, A.\ 2019, arXiv:1910.13970

\bibitem[Kardar(2007)]{kardar} Kardar, M.\ 2007, Cambridge University Press

\bibitem[Goldenfeld(1992)]{golden} Goldenfeld, N.\ 1992, Addison-Wesley

\bibitem[Planck Collaboration et al.(2020)]{Aghanim:2019ame} Planck Collaboration, Aghanim, N., Akrami, Y., et al.\ 2020, \aap, 641, A5. doi:10.1051/0004-6361/201936386

\bibitem[Planck Collaboration et al.(2020)]{Planck:2018lbu} Planck Collaboration, Aghanim, N., Akrami, Y., et al.\ 2020, \aap, 641, A8. doi:10.1051/0004-6361/201833886

\bibitem[Anderson et al.(2014)]{Anderson:2013zyy} Anderson, L., Aubourg, {\'E}., Bailey, S., et al.\ 2014, \mnras, 441, 24. doi:10.1093/mnras/stu523

\bibitem[Blake et al.(2011)]{wigglez} Blake, C., Kazin, E.~A., Beutler, F., et al.\ 2011, \mnras, 418, 1707. doi:10.1111/j.1365-2966.2011.19592.x

\bibitem[Ross et al.(2015)]{Ross:2014qpa} Ross, A.~J., Samushia, L., Howlett, C., et al.\ 2015, \mnras, 449, 835. doi:10.1093/mnras/stv154

\bibitem[Beutler et al.(2011)]{6df} Beutler, F., Blake, C., Colless, M., et al.\ 2011, \mnras, 416, 3017. doi:10.1111/j.1365-2966.2011.19250.x

\bibitem[Esteban et al.(2022)]{omena} Esteban, I., Mena, O., \& Salvado, J.\ 2022, arXiv:2202.04656
\end{thebibliography}

\end{document}